\documentstyle[12pt,axodraw]{article}

\begin{document}

\pagestyle{empty}

\begin{center} 
{\Large {\bf Proton decay in Supersymmetric 331 Model}}
\end{center}

\begin{center}
M. C. Rodriguez\footnote{mcrodriguez@fisica.furg.br} \\
{\it Funda\c c\~ao Universidade Federal do Rio Grande-FURG \\
Departamento de F\'\i sica \\
Av. It\'alia, km 8, Campus Carreiros \\
96201-900, Rio Grande, RS \\
Brazil}
\end{center}

\date{\today}


\begin{abstract}
The aim of this paper is to discuss  what sort of baryon-number 
violating process arise on the supersymmetric versions of the model based 
on the gauge group $SU(3)_{C}\otimes SU(3)_{L}\otimes U(1)_{N}$.  
We discuss the mechanism of baryon number violation which induces proton 
decay, and derive bounds on the relevant couplings.
\end{abstract}

\newpage

\section{Introduction}
\label{intro}

The Standard Model (SM)~\cite{sm} is exceedingly successful in describing leptons, 
quarks and their interactions. Nevertheless, the SM is 
not considered as the ultimate theory since neither the fundamental 
parameters, masses and couplings, nor the symmetry pattern are 
predicted. These elements are merely built into the model. Likewise, 
the spontaneous electroweak symmetry breaking is simply parametrized 
by a single Higgs doublet field. However, the necessity to go beyond it, from the 
experimental point of view, comes at the moment only from neutrino 
data~\cite{bk}. If neutrinos are massive then new physics beyond the SM is needed.

Theorists find supersymmetry appealing for reasons which are both
phenomenological and technical. The main phenomenological argument to study the Minimal Supersymmetric Standard Model 
(MSSM)~\cite{mssm} with soft-breaking terms ``at
the weak scale'' (i.e., with mass very roughly comparable to those of
the heaviest known elementary particles, the $W$ and $Z$ bosons and
the top quark) is that supersymmetric field theories ``naturally'' allow to choose the weak
scale to be many orders of magnitude below the hypothetical scale $\Lambda$
of Grand Unification or the Planck scale $M_{Pl}$. This is closely related to
the cancellation of quadratic divergencies \cite{2} in supersymmetric field
theories; such divergencies are notorious in non--supersymmetric theories with elementary 
scalar particles, such as the SM. The technical motivation is the idea that 
supersymmetry bring together bosons and fermions into generalized multiplets, 
others were attracted by the fact that local supersymmetry involves gravity.

In the MSSM theory if the top quark has an upper bound of 200 GeV the upper limit on the
mass of the lightest neutral scalar is less than $M_{Z}$ at the tree level \cite{japa}. 
This limit does not depend on the scale of supersymmetry breaking and holds independently of the short distance or large mass 
behavior of the theory \cite{kane}. This limit also does not depend on the number of the Higgs multiplet \cite{flores}.  
Radiative corrections rise the mass of the lightest neutral scalar to 130 GeV~\cite{haber2}. On another hand no direct 
observation of a Higgs boson has been made yet and current direct searches constraint its mass to $M_{h}>113$ GeV 
\cite{exp1} in the SM. 

On the other side, it is not clear what the physics beyond the SM should be. An interesting possibility is that at the TeV 
scale physics would be described by models which share some of the faults of the SM but give some insight concerning some 
questions which remain open in the SM context.

One of these  possibilities is that, at energies of a few TeVs, the gauge
symmetry may be $ SU(3)_c\otimes SU(3)_L \otimes U(1)_N$ (331) \cite{ppf,mpp}. The main motivations to study this kind of model are:
\begin{enumerate}
\item The family number must be three. This result comes from the fact that the model is anomaly-free only if we 
have equal number of triplets and antitriplets, counting the 
$SU(3)_{c}$ colors, and further more requiring the sum of all fermion charges 
to vanish. However each generation is anomalous, the anomaly 
cancellation occurs for the three, or multiply of three, together and not
generation by generation like in the SM. Therefore triangle anomalies together with asymptotic freedom imply that the number of generations 
must be three and only three. This may provides a first step towards answering the flavor question;
\item It explains why $\sin^{2} \theta_{W}<\frac{1}{4}$ is observed at the 
Z-pole.
This point come from the fact that in the model of Ref.~\cite{ppf} we have that
the $U(1)_N$ and $SU(3)_L$ coupling constants, $g'$ and $g$, respectively,
are related by  
\begin{equation}
t^{2}\equiv \left( \frac{g^{\prime}}{g} \right)^{2}= 
\frac{ \sin^{2} \theta_{W}}{1-4 \sin^{2} \theta_{W}}.
\label{eq2}
\end{equation}
Hence, this 331 model predicts that there exists an energy scale, 
say $\mu$, at which the model loses its perturbative character. 
The value of $\mu$ can be found through the condition   
$\sin^2\theta_W(\mu)=1/4$, and according to recent calculation 
$\mu \approx 4$ TeV \cite{fram,alex};
\item It is the simplest model that includes bileptons of both types: scalar 
and vectors ones. In fact, although there are several models which include doubly charged
scalar fields, not many of them incorporate doubly
charged vector bosons: this is a particularity of the 331 model
of Ref.~\cite{ppf};
\item The model has several sources of CP violation.
In the 331 model~\cite{ppf} we can implement the violation of the CP
symmetry, spontaneously~\cite{laplata1,dumm1} or explicitly~\cite{liung}. 
In models with exotic leptons it is possible to implement soft CP 
violation~\cite{cp3}; 
\item The extra neutral vector boson $Z^\prime$ conserves
flavor in the leptonic but not in the quark sector. The couplings to the leptons
are leptophobic because of the suppression factor $(1-4 \sin^{2} \theta_{W})^{1/2}$ but with 
some quarks there are enhancements because of the factor $(1-4 \sin^{2} \theta_{W})^{-1/2}$
~\cite{dumm2};
\item The charge quantization does not depend on the nature of the neutrino 
masses \cite{cp1};
\item The vectorial nature of the electromagnetic interaction is also 
explained \cite{cp2}.
\end{enumerate}
Recently, we have proposed the supersymmetric extensions of the 331 models 
\cite{331susy,331susy2}. In this kind of model the supersymmetry is naturally broken at 
a TeV scale, and the lightest scalar boson has an upper bound of 120 GeV.

Since it is possible to define the $R$-parity symmetry in both models, the phenomenology 
with $R$-parity conserved has similar features to that of the 
$R$-conserving MSSM: the supersymmetric particles are pair-produced and the
lightest neutralino is the lightest supersymmetric particle (LSP). 
However, there are differences between this kind of models and the MSSM with or without $R$-parity breaking.

In the case in the model of Ref.~\cite{331susy}, there are doubly charged scalar and vector fields. Hence, 
we have doubly charged charginos which are mixtures of the superpartners of the $U$-vector boson with the 
doubly charged scalars \cite{mcr}. This implies new interactions that are not present in the 
MSSM, for instance: $\tilde{\chi}^{--}\tilde{\chi}^0U^{++}$, $\tilde{\chi}^{-}\tilde{\chi}^{-}U^{++}$, 
$\tilde{l}^{-}l^{-}\tilde{\chi}^{++}$ where $\tilde{\chi}^{++}$ denotes any doubly charged chargino. 
Moreover, in the chargino production, besides the usual mechanism, we have additional contributions coming 
from the $U$-bilepton in the s-channel. Due to this fact we have an enhancement in the cross section of production 
of these particles in $e^{-}e^{-}$ collisors, such as the NLC~\cite{mcr}. 

We will also have the singly charged charginos 
and neutralinos, as in the MSSM, where there are processes like $\tilde{l}^{-}l^{+}\tilde{\chi}^{0}$, 
$\tilde{\nu_{L}}l^{-}\tilde{\chi}^{+}$, with $\tilde{l}$ denoting any slepton; $\tilde{\chi}^{-}$ denotes singly charged 
chargino and $\tilde{\nu}_L$ denotes any sneutrino. The only difference is that in the MSSM there are five neutralinos
while in our model there are eight neutralinos.

In the case of model in Ref.~\cite{331susy2} there is no double charged chargino. The mechanism of production 
of charginos and neutralinos are the same as in the MSSM, but in the case of $R$-parity conservation there are fifteen 
neutralinos and six singly charged chargino in this model.

However, $B$- and $L$-conservations are not ensured by gauge invariance and therefore it is worthwhile 
to investigate what happens when $R$-parity is violated. Remember that to give mass to neutrino in the MSSM we have 
to considerating $R$-parity violating interactions \cite{lepmass}. The $R$- parity violating terms will also induce proton decay. 
The decay of the proton is the most dramatic prediction coming from matter unification, and it put several constrain in biulding models. 

A general analysis of nucleon decay in supergravity unified models was presented in \cite{chamseddine}, where the authors included 
the full set of gaugino, gluino dressing diagrams in dimension five operators generated by Higgs-triplets exchange. These 
dimension five operators conserve $R$-parity, due it can not be forbidden in the theory, unless we use some discret symmetry to forbidden 
them. The problem with these operator generates the $p \rightarrow K^{+} \nu^{c}$ channel in the MSSM and in the minimal supersymmetric 
SU(5) model, it will be considered elsewhere \cite{alex1}. This channel is enough to exclude the minimal supersymmetric SU(5) model \cite{mura02}. Therefore, 
the proton 
decay is an important issue in any realistic extension of the SM. Here in this article, we are going to consider the mechanism that induce 
the proton decay in supersymmetric 331 models described above. 

This paper is organized as follows. In Section \ref{pmssm} we review the 
proton decay situation in the minimal supersymmetric standard model. In Section \ref{psusy1} we present the proton decay, this article is different than the 
approach used in \cite{pal}, in the minimal supersymmetric 331 model and  
in the subsequent section we present the result in supersymmetric 331 model with right-handed neutrinos. 
Our conclusions are found in the last section.

\section{Proton decay in the MSSM}
\label{pmssm}

In the MSSM~\cite{mssm}, the interactions are written in
terms of the left-handed (right-handed) $\hat{L}_{a}\sim({\bf2},-1)$ 
($\hat{l}^{c}_{a}\sim({\bf1},2)$) leptons, left-handed (right-handed) quarks 
$\hat{Q}_{i}\sim({\bf2},1/3)$ $(\hat{u}^{c}_{i}\sim({\bf1},-4/3),
\hat{d}^{c}_{i}\sim({\bf1},2/3))$; and the Higgs doublets
$\hat{H}_{1}\sim({\bf2},-1),\hat{H}_{2}\sim({\bf2},1)$. With those multiplets  
the superpotential that conserves $R$-parity is given by
$W_{2RC}+W_{3RC}+\overline{W}_{2RC}+\overline{W}_{3RC}$, where 
\begin{eqnarray}
W_{2RC}&=&\mu\epsilon\hat{H}_1\hat{H}_2,\nonumber \\ 
W_{3RC}&=&f^{l}_{ab}\epsilon\hat{L}_{a}\hat{H}_{1}\hat{l}^{c}_{b}+
f^{u}_{ij}\epsilon\hat{Q}_{i}\hat{H}_{2}\hat{u}^{c}_{j} +
f^{d}_{ij}\epsilon\hat{Q}_{i}\hat{H}_{1}\hat{d}^{c}_{j},
\label{mssmrpc}
\end{eqnarray}
while the $R$-parity violating terms are given by $W_{2RV}+W_{3RV}+
\overline{W}_{2RV}+\overline{W}_{3RV}$, where
\begin{eqnarray}
W_{2RV}&=&\mu_{0a}\epsilon \hat{L}_a\hat{H}_2,\nonumber \\
W_{3RV}&=&\lambda_{abc}\epsilon\hat{L}_{a}\hat{L}_{b}\hat{l}^{c}_{c}+
\lambda^{\prime}_{iaj}\epsilon\hat{Q}_{i}\hat{L}_{a}\hat{d}^{c}_{j}+ 
\lambda^{\prime\prime}_{ijk}\hat{d}^{c}_{i}\hat{u}^{c}_{j}\hat{d}^{c}_{k},
\label{mssmrpv}
\end{eqnarray}
and we have suppressed $SU(2)$ indices, $\epsilon$ is the
antisymmetric $SU(2)$ tensor. Above, and below in the following, the subindices
$a,b,c$ run over the lepton generations $e,\mu,\tau$ but a superscript $^c$
indicates charge conjugation; $i,j,k=1,2,3$ denote quark generations.

The couplings $\lambda$ and $\lambda^{\prime}$ types are $L$-violating
while $\lambda^{\prime \prime}$ types are $B$-violating Yukawa couplings.
$\lambda_{abc}$ is antisymmetric under the interchange of the first
two generation indices, while $\lambda^{\prime \prime}_{ijk}$ is antisymmetric under
the interchange of $i$ and $k$ indices. Eq.(\ref{mssmrpv}) thus contains 27
$\lambda'$-type and 9 each of $\lambda$- and $\lambda''$-type
couplings. Hence including the 3 additional bilinear $\mu_{0}$-terms, there are $9+27+9+3=48$ 
new terms beyond those of the MSSM.

From the Eq.(\ref{mssmrpv}) we can get the followings interaction, from the 
$B$-violating Yukawa couplings
\begin{eqnarray}
\lambda^{\prime \prime}_{ijk} \bar{d}_{iR}u^{c}_{jL}\tilde{d}^{c}_{k}+H.c. \,\ .
\label{lpp}
\end{eqnarray}
and from $\lambda^{\prime}_{iaj}$ we get
\begin{eqnarray}
\lambda^{\prime}_{iaj}\left( \bar{d}^{c}_{iR}\nu_{aL}- \bar{u}^{c}_{iR}l_{aL} \right) \tilde{d}_{j}+H.c. \,\ .
\label{lpmssm}
\end{eqnarray}
From the Eqs.(\ref{lpp},\ref{lpmssm}) we can draw two Feynmann diagrams 
to the proton decay, shown in Figs.(1,2). 

On the first figure, Fig.(1), we have the proton decay on charged leptons. 
Considerating, as first approximation, no mixing in the quarks, neutrinos and squarks sectors, it means 
$u_{1} \equiv u$, $u_{2} \equiv c$ and $u_{3} \equiv t$ and so on. The proton could decay 
in $p \to \pi^{0}e^{+}$, $p \to \bar{D}^{0} \mu^{+}$ and $p \to t^{c} \tau^{+}$, 
but the last two contribution are forbidden kinematically.
\begin{figure}[t]
%
%
\begin{center}
\begin{picture}(150,100)(0,0)
\SetWidth{1.2}
\ArrowLine(10,10)(50,50)
\Text(10,10)[r]{$u_{1}$}
\ArrowLine(10,90)(50,50)
\Text(10,90)[r]{$d_{1}$}
\Vertex(50,50){2.0}
\Text(30,50)[]{$\lambda^{\prime \prime}$}
\DashLine(100,50)(50,50){3}
\Vertex(100,50){2.0}
\Text(115,50)[]{$\lambda^{\prime}$}
\Text(75,58)[b]{$\tilde{d}^{c}_{2},\tilde{d}^{c}_{3}$}
\ArrowLine(140,90)(100,50)
\Text(145,90)[l]{$u^{c}_{1},u^{c}_{2},u^{c}_{3}$}
\ArrowLine(140,10)(100,50)
\Text(145,10)[l]{$l^{+}_{1},l^{+}_{2},l^{+}_{3}$}
\end{picture}\\ 
{\sl Figure 1: Proton Decay in charged leptons in the MSSM.}
\end{center}
\label{f1}
\end{figure}
Due the analysis presented above the proton can decay only in 
$p \to \pi^{0}e^{+}$. On  dimensional grounds we estimate 
\begin{eqnarray}
\Gamma (p \to \pi^{0} e^{+})\approx \frac{\alpha(\lambda^{\prime}_{11k})
\alpha(\lambda^{\prime \prime}_{11k})}{{\tilde m}_{d_{k}}^4}{M_{proton}^5}.
\end{eqnarray}
Here $\alpha(\lambda)=\lambda^2/(4\pi)$. Given that $\tau(p \to e \pi)
>1.6 \times 10^{33}$ years \cite{pdg} and considerating 
$\tilde{m}_{d_{k}} \sim O(1TeV)$, we obtain 
\begin{equation}
\lambda^{\prime}_{11k}\lambda^{\prime \prime}_{11k}\; < \; 5.29 \times 
10^{-26}. 
\label{proton}
\end{equation}
It is consistent with the limits presented in \cite{drei}. For a more detailed 
calculation see \cite{sher,hin}. There are another decay mode, where the 
proton decay in antineutrino, it was considered in \cite{smi}. In this case, 
we get the same numerical results as presented in Eq.(\ref{proton}). 
\begin{figure}[t]
%
%
\begin{center}
\begin{picture}(150,100)(0,0)
\SetWidth{1.2}
\ArrowLine(10,10)(50,50)
\Text(10,10)[r]{$u_{1}$}
\ArrowLine(10,90)(50,50)
\Text(10,90)[r]{$d_{1}$}
\Vertex(50,50){2.0}
\Text(30,50)[]{$\lambda^{\prime \prime}$}
\DashLine(100,50)(50,50){3}
\Vertex(100,50){2.0}
\Text(115,50)[]{$\lambda^{\prime}$}
\Text(75,58)[b]{$\tilde{d}^{c}_{2},\tilde{d}^{c}_{3}$}
\ArrowLine(140,90)(100,50)
\Text(145,90)[l]{$d^{c}_{1},d^{c}_{2},d^{c}_{3}$}
\ArrowLine(140,10)(100,50)
\Text(145,10)[l]{$\nu^{c}_{1},\nu^{c}_{2},\nu^{c}_{3}$}
\end{picture}\\ 
{\sl Figure 2: Proton Decay in antineutrinos in the  MSSM.}
\end{center}
\label{f2} 
\end{figure}

The bound presented in Eq.(\ref{proton}) is so strict that the 
only natural explanation is for at least one of the couplings to be 
zero.Thus the simplest supersymmetric extension of the standard model is excluded.

The baryon-parity is defined as
\begin{eqnarray}
( \hat{Q}_{i},\hat{u}^{c}_{i},\hat{d}^{c}_{i}) \rightarrow - ( \hat{Q}_{i},\hat{u}^{c}_{i},\hat{d}^{c}_{i}),\nonumber \\
( \hat{L}_{a},\hat{l}^{c}_{a},\hat{H}_{1},\hat{H}_{2}) \rightarrow ( \hat{L}_{a},\hat{l}^{c}_{a},\hat{H}_{1},\hat{H}_{2}).
\label{eureka}
\end{eqnarray}
This symmetry forbid the proton decay in the MSSM, because it allow only interactions that violates $L$-number, therefore the proton 
is stable at tree-level. It is interesting using this symmetry, because 
Leptogenesis \cite{fy86,lep} provides a 
simple and elegant explanation of the cosmological matter-antimatter asymmetry. There is a large number of discrete symmetries 
which can avoid the proton decay \cite{drei}.   

If you want to keep the proton decay it 
is necessary to consider the effects coming from the heavier quarks and squarks sectors. How we neglected the intergerenacional mixing 
ones can ask: What are the bounds on the 
couplings involving heavy generations? Whether there are unrestricted couplings? These 
questions are  important and in \cite {smi} the authors proved that operators relevant 
to the proton decay are always present in the one loop effective lagrangian, and they imply strong bounds on any product 
of the couplings $\lambda^{\prime}\lambda^{\prime \prime}$. For squark masses below 1 TeV 
they found the conservative bounds $|\lambda^{\prime}\lambda^{\prime \prime}| \le 10^{-9}$ in absence of squark flavor mixing, 
and $|\lambda^{\prime}\lambda^{\prime \prime}| \le 10^{-11}$ when this mixing is taken into account. 
We have to enphasasive that in this reference they neglected the mixing in the neutrino's sector.

\section{Proton decay in the Minimal Supersymmetric 331 Model}
\label{psusy1}

In the nonsupersymmetric 331 model~\cite{ppf} the fermionic representation
content is as follows: left-handed leptons $L_{a}=(\nu_{a},l_{a},l^{c}_{a})_L
\sim({\bf1},{\bf3},0)$, $a=e,\mu,\tau$; left-handed quarks 
$Q_{\alpha L}=(d_{\alpha},u_{\alpha},j_{\alpha}) \sim({\bf3},{\bf3}^*,-1/3)$, $\alpha=1,2$, 
$Q_{3L}=(u_{3},d_{3},J)\sim({\bf3},{\bf3},2/3)$; and in the right-handed components we have 
$u^{c}_{i},d^{c}_{i},\,i=1,2,3$, that transform as in the standard model, and the exotic quarks 
$j^{c}_{\alpha}\sim({\bf3}^{*},{\bf1},4/3),J^{c}\sim({\bf3}^{*},{\bf1},-5/3)$.
The minimal scalar representation content is formed by three scalar triplets: 
$\eta\sim({\bf1},{\bf3},0)=(\eta^{0},\eta^{-}_{1},\eta^{+}_{2})^T$;
$\rho\sim({\bf1},{\bf3},+1)=(\rho^{+}, \rho^{0}, \rho^{++})^T$ and 
$\chi\sim({\bf1},{\bf3},-1)=(\chi^{-},\chi^{--},\chi^{0})^T$, and one scalar antisextet
$S\sim({\bf1},{\bf6}^{*},0)$.We can avoid the introduction of the antisextet by adding a charged lepton
transforming as a singlet~\cite{dma,lepmass1}. Notwithstanding, here we will
omit both the antisextet and the exotic lepton, because we have showed in \cite{lepmass} that in this model the
$R$-violating interactions give the correct masses to $e,\mu$ and $\tau$. The complete set of fields in the 
minimal supersymmetric 331 model(MSUSY331) has been given in \cite{331susy,mcr}.

In this model the superpotential is given by
\begin{equation}
W=W_{2}+W_{3}+\overline{W}_{2}+\overline{W}_{3},
\label{sp331}
\end{equation}
where
\begin{eqnarray}
W_{2}&=&\mu_{0a}\hat{L}_{a} \hat{ \eta}^{\prime}+ 
\mu_{ \eta} \hat{ \eta} \hat{ \eta}^{\prime}+
 \mu_{ \rho} \hat{ \rho} \hat{ \rho}^{\prime}+ 
\mu_{ \chi} \hat{ \chi} \hat{ \chi}^{\prime}, \nonumber \\
W_{3}&=& \lambda_{1abc} \epsilon \hat{L}_{a} \hat{L}_{b} \hat{L}_{c}+
\lambda_{2ab} \epsilon \hat{L}_{a} \hat{L}_{b} \hat{ \eta}+
\lambda_{3a} \epsilon \hat{L}_{a} \hat{\chi} \hat{\rho}+
f_{1} \epsilon \hat{ \rho} \hat{ \chi} \hat{ \eta}+ 
f^{\prime}_{1}\epsilon \hat{ \rho}^{\prime}\hat{ \chi}^{\prime}\hat{ \eta}^{\prime} 
\nonumber \\
&+&  \kappa_{1i} \hat{Q}_{3} \hat{\eta}^{\prime} \hat{u}^{c}_{i}
+\kappa_{2i} \hat{Q}_{3} \hat{\rho}^{\prime} \hat{d}^{c}_{i}+
\kappa_{3} \hat{Q}_{3} \hat{\chi}^{\prime} \hat{J}^{c}+
\kappa_{4\alpha i} \hat{Q}_{\alpha} \hat{\eta} \hat{d}^{c}_{i}+
\kappa_{5\alpha i} \hat{Q}_{\alpha} \hat{\rho} \hat{u}^{c}_{i} +
\kappa_{6\alpha \beta} \hat{Q}_{\alpha} \hat{\chi} \hat{j}^{c}_{\beta} \nonumber \\
&+& 
\lambda^{\prime}_{\alpha ai} \hat{Q}_{\alpha} \hat{L}_{a} \hat{d}^{c}_{i}+
\lambda^{\prime \prime}_{ijk} \hat{d}^{c}_{i}\hat{u}^{c}_{j} \hat{d}^{c}_{k}+
\xi_{1ij \beta} \hat{u}^{c}_{i} \hat{u}^{c}_{j} \hat{j}^{c}_{\beta}+
\xi_{2i \beta} \hat{d}^{c}_{i} \hat{J}^{c} \hat{j}^{c}_{\beta}. \nonumber \\
\label{sp}
\end{eqnarray}
The term proportional to $\mu_{0}$ and the terms in the last line in the equation above, they don't conserve $R$ parity \cite{331susy}.
The term $\lambda_{1}$ is totally antisymmetric in the generation indices. So, for three generations, there is only one such
independent coupling, which we will denote simply by $\lambda$ \cite{pal}. The terms $\mu_{0}$,$\lambda^{\prime}$, $\xi_{1}$ and 
$\xi_{2}$ are L-violating terms, while $\lambda^{\prime \prime}$ don't conserve baryon number. The $\lambda^{\prime}$ term 
contribute with 12 aditional parameters; 
while we have 9 $\lambda^{\prime \prime}$-type; 1 $\lambda_{1}$-type; 
6 each of $\xi_{1}$-type and $\xi_{2}$-type couplings and 3 $\mu_{0}$ 
parameter. Therefore, there are 37 
aditional parameters in relation with the model that conserve $R$ parity. We shall comment that the terms $\xi_{1ij \beta}$ 
and $\xi_{2i \beta}$ involve exotic quark fields, and can not affect processes involving the usual quarks at the tree level.
 
The term $\lambda^{\prime \prime}$ is present in the MSSM, see Eq.(\ref{mssmrpv}).
Due this, fact, this term, gives the same interaction as in Eq.(\ref{lpp}). 
The $\lambda^{\prime}$ term gives the followings interactions
\begin{equation}
\lambda^{\prime}_{\alpha ai}\left( \bar{d}^{c}_{\alpha R}\nu_{aL}+ \bar{u}^{c}_{\alpha R}l_{aL} \right) \tilde{d}^{c}_{i}+H.c. \,\ .
\label{lp331}
\end{equation}
We have to emphasize that this contribution is different from that presented 
in Eq.(\ref{lpmssm}). 
The difference between the two equation is the following: the outcoming quark 
in Eq.(\ref{lp331}) can belong only to the first or the second generation, remember that $\alpha =1,2$, although 
in Eq.(\ref{lpmssm}) came from any of the three family, $i=1,2,3$. Here our notation is not the same as in 
\cite{pal}, where the outcoming quark in Eq.(\ref{lp331}) can belong only to the second or the third generation.

Which the vertices given in the Eqs.(\ref{lpp},\ref{lp331}), we have similar diagrams as presented in 
the MSSM case, the diagrams in this model are drawn in Figs.(3,4).
\begin{figure}[t]
%
%
\begin{center}
\begin{picture}(150,100)(0,0)
\SetWidth{1.2}
\ArrowLine(10,10)(50,50)
\Text(10,10)[r]{$u_{1}$}
\ArrowLine(10,90)(50,50)
\Text(10,90)[r]{$d_{1}$}
\Vertex(50,50){2.0}
\Text(30,50)[]{$\lambda^{\prime \prime}$}
\DashLine(100,50)(50,50){3}
\Vertex(100,50){2.0}
\Text(115,50)[]{$\lambda^{\prime}$}
\Text(75,58)[b]{$\tilde{d}^{c}_{2},\tilde{d}^{c}_{3}$}
\ArrowLine(140,90)(100,50)
\Text(145,90)[l]{$u^{c}_{1},u^{c}_{2}$}
\ArrowLine(140,10)(100,50)
\Text(145,10)[l]{$l^{+}_{1},l^{+}_{2},l^{+}_{3}$}
\end{picture}\\ 
{\sl Figure 3: Proton Decay in charged leptons in the MSUSY331.}
\end{center}
\label{f3} 
\end{figure}
Barring contributions coming from intergenerational mixing, as first approximation, the Fig.(3) 
would produce the following dominant decay is
\begin{eqnarray}
p \to \pi^{0} e^{+}.
\end{eqnarray}

\begin{figure}[t]
%
%
\begin{center}
\begin{picture}(150,100)(0,0)
\SetWidth{1.2}
\ArrowLine(10,10)(50,50)
\Text(10,10)[r]{$u_{1}$}
\ArrowLine(10,90)(50,50)
\Text(10,90)[r]{$d_{1}$}
\Vertex(50,50){2.0}
\Text(30,50)[]{$\lambda^{\prime \prime}$}
\DashLine(100,50)(50,50){3}
\Vertex(100,50){2.0}
\Text(115,50)[]{$\lambda^{\prime}$}
\Text(75,58)[b]{$\tilde{d}^{c}_{2},\tilde{d}^{c}_{3}$}
\ArrowLine(140,90)(100,50)
\Text(145,90)[l]{$d^{c}_{1},d^{c}_{2}$}
\ArrowLine(140,10)(100,50)
\Text(145,10)[l]{$\nu^{c}_{1},\nu^{c}_{2},\nu^{c}_{3}$}
\end{picture}\\ 
{\sl Figure 4: Proton Decay in antineutrinos in the MSUSY331.}
\end{center}
\label{f4} 
\end{figure}
In the case of Fig.(4) the dominant decay mode would be
\begin{eqnarray}
p \to \pi^{+} \nu^{c}_{e} \,,
\end{eqnarray}
this will lead to a lifetime of 
\begin{eqnarray}
\tau(p \to \pi^{+}\nu^{c}_{e}) \simeq \left( \frac{(4 \pi)^{2} \tilde{m}^{4}_{d_{k}} }
{(\lambda^{\prime}_{11k}\lambda^{\prime \prime}_{11k})^{2}M_{proton}^{5}} \right) \, .
\label{pksusy331} 
\end{eqnarray} 
Using the same parameters as discussed above Eq.(\ref{proton}), we obtain the same 
bound as got in the MSSM case.  

Of course, intergenerational mixings coming from the neutrino's sector and 
from the $d$-quark and $\tilde{d}$-squarks sectors  will suppress 
decays like $p\to \pi^{+}\nu_{e}$, as discussed in \cite{smi}, and therefore
the bounds given in Eq.(\ref{proton}) will be a little weaker. We don't perform this calculation here, because 
radiative corrections have to be taken into account, however, this 
has to be done in the context of the MSUSY331 model which is not in the scope of the present work.

Let us now look at a different mechanism for nucleon decay. Due the fact that 
in the model there is an ineraction as ${\cal L}_{l \tilde{l} \tilde{V}}$ and ${\cal L}_{q \tilde{q} \tilde{V}}$ \cite{mcr}
toghether with Eqs.(\ref{lpp},\ref{lp331}) results in a Feynmann diagram given by Fig.(5).
\begin{figure}[t]
%
%
\begin{center}
\begin{picture}(220,100)(0,0)
\SetWidth{1.1}
\ArrowLine(10,10)(50,50)
\Text(10,10)[r]{$u_{1}$}
\ArrowLine(10,90)(50,50)
\Text(10,90)[r]{$d_{1}$}
\Vertex(50,50){2.0}
\Text(30,50)[]{$\lambda^{\prime \prime}$}
\DashLine(90,50)(50,50){3}
\Vertex(90,50){2.0}
\Text(80,40)[]{$g$}
\Text(70,55)[b]{$\tilde{d}^{c}_{2}$}
\ArrowLine(90,90)(90,50)
\Text(95,90)[l]{$d^{c}_{2}$}
\Line(90,50)(130,50)
\Text(110,45)[t]{$\tilde{\gamma}, \tilde{Z}, \tilde{Z}^{\prime}$}
\Vertex(130,50){2.0}
\Text(140,60)[]{$g$}
\ArrowLine(130,50)(130,90)
\Text(160,90)[b]{$l^{-}_{1},l^{-}_{2},l^{-}_{3}$}
\DashLine(170,50)(130,50){3}
\Vertex(170,50){2.0}
\Text(190,50)[]{$\lambda$}
\Text(160,30)[b]{$\tilde{l}^{+}_{1},\tilde{l}^{+}_{2},\tilde{l}^{+}_{3}$}
\ArrowLine(170,50)(210,90)
\Text(220,90)[l]{$\nu_{1},\nu_{2},\nu_{3}$}
\ArrowLine(210,10)(170,50)
\Text(220,10)[l]{$l^{+}_{1},l^{+}_{2},l^{+}_{3}$}
\end{picture}\\ 
{\sl Figure 5: Proton Decay in two charged leptons and one neutrino and one meson in antineutrinos in the MSUSY331.}
\end{center}
\label{f5}
\end{figure}
In this case, the three outgoing leptonic
fields belong to three different generations. Since the $\tau$-lepton
is heavier than the proton, this means that the charged leptons
available in the decay product must be $e^{-}\mu^{+}$ or
$e^{+}\mu^{-}$. In other words, we obtain the following decay modes
at the quark level:
\begin{eqnarray}
u d \to s^{c}e^{\mp}\mu^{\pm}\nu_{\tau} \,,
\end{eqnarray}
taking into account that the couplings $\lambda^{\prime \prime}$ must be
antisymmetric in the down-type quark indices.
For the proton, it implies the decay mode
\begin{eqnarray}
p \to K^{+}e^{\mp}\mu^{\pm}\nu_{\tau} \,.
\end{eqnarray}
This decay mode have never been observed yet; experimenters are urged to examine this proton mode decay. This process 
will give a lifetime \cite{pal}
\begin{eqnarray}
\tau(p \to K^{+}e^{\mp}\mu^{\pm}\nu_{\tau}) \simeq 
\left( \frac{(4 \pi)^{2} \tilde{m}^{4}_{d_{2}}\tilde{m}^{4}_{l_{a}}\tilde{m}^{2}_{Z}}
{g^{4}(\lambda\lambda^{\prime \prime}_{112})^{2}M_{proton}^{11}}\right)\,. 
\end{eqnarray}
Using that the bound in proton mean life is $\tau>1.6 \times 10^{25}$ years \cite{pdg} and assuming the superpartner 
masses in the range of 1 TeV, we can found the following limit
\begin{equation}
\lambda\lambda^{\prime \prime}_{112}<1.51 \times 10^{-12}.
\label{proton2}
\end{equation}  

Of course, we can also have $p \to \pi^{+} e^{\mp}\mu^{\pm}\nu_{\tau}$ etc, but those will be suppressed by intergenerational 
mixings, as mentioned, in the previous discussion.

The baryon-parity forbid the terms in the last line in the Eq.(\ref{sp}), except the 
term $\lambda^{\prime}_{\alpha ai}$, due the fact that it contain only two quarks superfields. Therefore the proton is stable at 
tree-level, altought it is possible to have Leptogenesis and to produce 
massive neutrinos. 

\section{Proton decay in in the Minimal Supersymmetric 331 Model with right-handed neutrinos}
\label{psusy2}

In the nonsupersymmetric 331 model with right-handed neutrinos 
~\cite{mpp} the fermionic representation
content is as follows: left-handed leptons and quarks $L_{a}=(\nu_{a},l_{a}, \nu^{c}_{a})_L
\sim({\bf1},{\bf3},-1/3)$, $a=e,\mu,\tau$; $Q_{\alpha L}=(d_{\alpha},u_{\alpha},d^{\prime}_{\alpha}) 
\sim({\bf3},{\bf3}^*,0)$, $\alpha=1,2$, 
$Q_{3L}=(u_{3},d_{3},u^{\prime})\sim({\bf3},{\bf3},1/3)$; and in the right-handed components we have 
$l^{c}_{a}\sim({\bf1},{\bf1},1)$, $u^{c}_{i},d^{c}_{i},\,i=1,2,3$, that transform as in the standard model, and 
the exotic quarks $u^{\prime c}\sim({\bf3}^*,{\bf1},-2/3),d^{\prime c}_{\alpha}\sim({\bf3}^*,{\bf1},1/3)$.
The minimal scalar representation content is formed by three scalar triplets: 
$\eta\sim({\bf1},{\bf3},-1/3)=(\eta^{0}_{1},\eta^{-},\eta^{0}_{2})^T$; 
$\chi\sim({\bf1},{\bf3},-1/3)=(\chi^{0}_{1},\chi^{-},\chi^{0}_{2})^T$ and  
$\rho\sim({\bf1},{\bf3},2/3)=(\rho^{+}_{1}, \rho^{0},\rho^{+}_{2})^T$. The complet set of fields 
on this supersymmetric 331 model (331SUSYRN) is given in \cite{331susy2}.

In this model the superpotencial is given again by Eq.(\ref{sp331}), where
\begin{eqnarray}
W_{2}&=&\mu_{0a}\hat{L}_{a} \hat{ \eta}^{\prime}+
\mu_{1a}\hat{L}_{a} \hat{ \chi}^{\prime}+ 
\mu_{ \eta} \hat{ \eta} \hat{ \eta}^{\prime}+
\mu_{ \chi} \hat{ \chi} \hat{ \chi}^{\prime}+
 \mu_{ \rho} \hat{ \rho} \hat{ \rho}^{\prime}, \nonumber \\
W_{3}&=& \lambda_{1ab} \hat{L}_{a} \hat{ \rho}^{\prime} \hat{l}^{c}_{b}+ 
\lambda_{2a} \epsilon \hat{L}_{a} \hat{\chi} \hat{\rho}+
\lambda_{3a} \epsilon \hat{L}_{a} \hat{\eta} \hat{\rho}+
\lambda_{4ab} \epsilon \hat{L}_{a} \hat{L}_{b} \hat{\rho}+
\kappa_{1i} \hat{Q}_{3} \hat{\eta}^{\prime} \hat{u}^{c}_{i}+
\kappa_{1}^{\prime} \hat{Q}_{3} \hat{\eta}^{\prime} \hat{u}^{\prime c} \nonumber \\ 
&+&
\kappa_{2i} \hat{Q}_{3} \hat{\chi}^{\prime} \hat{u}^{c}_{i}+
\kappa_{2}^{\prime} \hat{Q}_{3} \hat{\chi}^{\prime} \hat{u}^{\prime c}+
\kappa_{3\alpha i} \hat{Q}_{\alpha} \hat{\eta} \hat{d}^{c}_{i}+
\kappa_{3\alpha \beta}^{\prime} \hat{Q}_{\alpha} \hat{\eta} \hat{d}^{\prime c}_{\beta}+
\kappa_{4\alpha i} \hat{Q}_{\alpha}\hat{\rho}\hat{u}^{c}_{i}+
\kappa_{4\alpha}^{\prime} \hat{Q}_{\alpha}\hat{\rho}\hat{u}^{\prime c} \nonumber \\
&+&
\kappa_{5i}\hat{Q}_{3} \hat{\rho}^{\prime} \hat{d}^{c}_{i}+
\kappa_{5 \beta}^{\prime}\hat{Q}_{3} \hat{\rho}^{\prime} \hat{d}^{c}_{\beta}+
\kappa_{6\alpha i} \hat{Q}_{\alpha} \hat{\chi} \hat{d}^{c}_{i}+
\kappa_{6\alpha \beta}^{\prime} \hat{Q}_{\alpha} \hat{\chi} \hat{d}^{\prime c}_{\beta}+
f_{1} \epsilon \hat{ \rho} \hat{ \chi} \hat{ \eta}+
f^{\prime}_{1} \epsilon \hat{ \rho}^{\prime}\hat{ \chi}^{\prime}\hat{ \eta}^{\prime} \nonumber \\
&+&
\zeta_{\alpha \beta \gamma} \epsilon \hat{Q}_{\alpha} \hat{Q}_{\beta} \hat{Q}_{\gamma}+ 
\lambda^{\prime}_{\alpha ai}\hat{Q}_{\alpha}\hat{L}_{a} \hat{d}^{c}_{i}+
\lambda^{\prime \prime}_{ijk} \hat{d}^{c}_{i} \hat{u}^{c}_{j} \hat{d}^{c}_{k}+
\xi_{1ij \beta} \hat{d}^{c}_{i} \hat{u}^{c}_{j} \hat{d}^{\prime c}_{\beta}+
\xi_{2 \alpha a \beta}\hat{Q}_{\alpha} \hat{L}_{a} \hat{d}^{\prime c}_{\beta} \nonumber \\
&+&
\xi_{3i \beta} \hat{d}^{c}_{i} \hat{u}^{\prime c} \hat{d}^{\prime c}_{\beta}+
\xi_{4ij} \hat{d}^{c}_{i} \hat{u}^{\prime c} \hat{d}^{c}_{j}+
\xi_{5 \alpha i \beta} \hat{d}^{\prime c}_{\alpha} \hat{u}^{c}_{i} \hat{d}^{\prime c}_{\beta}+
\xi_{6 \alpha \beta} \hat{d}^{\prime c}_{\alpha}\hat{u}^{\prime c} \hat{d}^{\prime c}_{\beta} 
\label{sp3susy2}
\end{eqnarray}
The terms proportional to $\mu_{0a}$, $\mu_{1a}$ and the last two lines in the equation above don't conserve $R$ parity. 
The terms $\mu_{0}$,$\mu_{1}$,$\lambda^{\prime}$ and $\xi_{2}$ are L-violating terms, 
while $\zeta$, $\lambda^{\prime \prime}$, $\xi_{1}$, $\xi_{3}$, $\xi_{4}$, $\xi_{5}$ and $\xi_{6}$ 
don't conserve baryon number. The terms $\mu_{0}$, $\mu_{1}$ contribute each with 
3 aditional parameter. The term $\zeta_{\alpha \beta \gamma}$ is totally 
antisymmetric in the generation indices. So, for three generations, there is only one such independent coupling, which we 
will denote simply by $\zeta$.  There are $\lambda^{\prime}$ there are 12 aditional parameters; while we have 9 $\lambda^{\prime \prime}$-type;18 $\xi_{1}$
-type;12 $\xi_{2}$-type;6 $\xi_{3}$-type;3 $\xi_{4}$-type;3 $\xi_{5}$-type and 
1 $\xi_{6}$-type coupling. In this model, there are 71 aditional parameters in relation with the model with conserve $R$ parity. We shall comment 
that the terms $\xi_{1ij \beta}$, $\xi_{2 \alpha a \beta}$,  $\xi_{3i \beta}$, $\xi_{4ij}$, $\xi_{5 \alpha i \beta}$ and 
$\xi_{6 \alpha \beta}$ involve exotic quark fields, and can not affect processes involving the usual quarks at the tree level.

The terms $\lambda^{\prime}$ and $\lambda^{\prime \prime}$ are the same as the last model, see Eq.(\ref{sp}), 
and they will give the same vertices as shown in Eqs.(\ref{lpp},\ref{lp331}).

The $\zeta$ term gives the followings interactions
\begin{equation}
\zeta[ \bar{d}^{c}_{\alpha R}u_{\beta L}\tilde{d}^{\prime}_{\gamma}+
\bar{d}^{\prime c}_{\beta R}u_{\alpha L}\tilde{d}_{\gamma}]+H.c. \,\ .
\label{z3312}
\end{equation}

The terms $\xi_{1ij \beta}$ and $\xi_{2 \alpha a \beta}$ gives the followings interactions
\begin{eqnarray}
&\xi_{1ij \beta}& \bar{d}_{iR}u^{c}_{jL}\tilde{d}^{\prime c}_{\beta}+H.c. \,\ , \nonumber \\
&\xi_{2 \alpha a \beta}& [ (\bar{d}^{c}_{\alpha R}\nu_{aL}+ \bar{u}^{c}_{\alpha R}l_{aL})\tilde{d}^{\prime c}_{\beta}]+H.c. \,\ .
\label{dtilde}
\end{eqnarray}

Wchich the vertices given in the Eqs.(\ref{lpp},\ref{lp331}), we have similar diagrams as in 
previous section. However the Eq.(\ref{dtilde}) give two more contributions drawn in 
Figs.(6,7)
\begin{figure}[t]
%
%
\begin{center}
\begin{picture}(150,100)(0,0)
\SetWidth{1.2}
\ArrowLine(10,10)(50,50)
\Text(10,10)[r]{$u_{1}$}
\ArrowLine(10,90)(50,50)
\Text(10,90)[r]{$d_{1}$}
\Vertex(50,50){2.0}
\Text(30,50)[]{$\xi_{1}$}
\DashLine(100,50)(50,50){3}
\Vertex(100,50){2.0}
\Text(115,50)[]{$\xi_{2}$}
\Text(75,58)[b]{$\tilde{d}^{\prime c}_{1},\tilde{d}^{\prime c}_{2}$}
\ArrowLine(140,90)(100,50)
\Text(145,90)[l]{$u^{c}_{1},u^{c}_{2}$}
\ArrowLine(140,10)(100,50)
\Text(145,10)[l]{$l^{+}_{1},l^{+}_{2},l^{+}_{3}$}
\end{picture}\\ 
{\sl Figure 6: Proton Decay via $d^{\prime}$ to charged leptons in the 331SUSYRN.}
\end{center}
\label{f6}
\end{figure}

\begin{figure}[t]
%
%
\begin{center}
\begin{picture}(150,100)(0,0)
\SetWidth{1.2}
\ArrowLine(10,10)(50,50)
\Text(10,10)[r]{$u_{1}$}
\ArrowLine(10,90)(50,50)
\Text(10,90)[r]{$d_{1}$}
\Vertex(50,50){2.0}
\Text(30,50)[]{$\xi_{1}$}
\DashLine(100,50)(50,50){3}
\Vertex(100,50){2.0}
\Text(115,50)[]{$\xi_{2}$}
\Text(75,58)[b]{$\tilde{d}^{\prime c}_{1},\tilde{d}^{\prime c}_{2}$}
\ArrowLine(140,90)(100,50)
\Text(145,90)[l]{$d^{c}_{1},d^{c}_{2}$}
\ArrowLine(140,10)(100,50)
\Text(145,10)[l]{$\nu^{c}_{1},\nu^{c}_{2},\nu^{c}_{3}$}
\end{picture}\\ 
{\sl Figure 7: Proton Decay $d^{\prime}$ to neutral leptons in the 331SUSYRN.}
\end{center}
\label{f7}
\end{figure}
Remember that in the MSSM $\tilde{m}_{d} \propto m_{d}$, and due the fact that 
$m_{d^{\prime}}>m_{d}$ then it implies that $\tilde{m}_{{d}^{\prime}}>\tilde{m}_{d}$. Due this 
fact we can negligible this extras contribution, and the get the same answer as given in 
Eq.(\ref{pksusy331}), with the same bounds presented in Eq.(\ref{proton}).

On this model, there are, another contribution to the proton decay, as shown in Fig.(5), 
but in this model the right vertex is changed to the vertex shown in Eq.(\ref{z3312}). The 
Feynmann diagrams on this model are shown in Figs.(8,9) but we see that in this 
case, there are exotic quarks at the final state, therefore this process are forbiden kinematically.
\begin{figure}[t]
%
%
\begin{center}
\begin{picture}(220,100)(0,0)
\SetWidth{1.1}
\ArrowLine(10,10)(50,50)
\Text(10,10)[r]{$u_{1}$}
\ArrowLine(10,90)(50,50)
\Text(10,90)[r]{$d_{1}$}
\Vertex(50,50){2.0}
\Text(30,50)[]{$\lambda^{\prime \prime}$}
\DashLine(90,50)(50,50){3}
\Vertex(90,50){2.0}
\Text(80,40)[]{$g$}
\Text(70,55)[b]{$\tilde{d}_{2}^{c}$}
\ArrowLine(90,90)(90,50)
\Text(95,90)[l]{$d^{c}_{2}$}
\Line(90,50)(130,50)
\Text(110,45)[t]{$\tilde{\gamma}, \tilde{Z}, \tilde{Z}^{\prime}$}
\Vertex(130,50){2.0}
\Text(140,60)[]{$g$}
\ArrowLine(130,50)(130,90)
\Text(160,90)[b]{$d_{1},d_{2}$}
\DashLine(170,50)(130,50){3}
\Vertex(170,50){2.0}
\Text(190,50)[]{$\zeta$}
\Text(160,30)[b]{$\tilde{d}^{c}_{1},\tilde{d}^{c}_{2}$}
\ArrowLine(210,90)(170,50)
\Text(220,90)[l]{$u^{c}_{1},u^{c}_{2}$}
\ArrowLine(210,10)(170,50)
\Text(220,10)[l]{$d^{\prime c}_{1},d^{\prime c}_{2}$}
\end{picture}\\ 
{\sl Figure 8: Proton Decay in exotic quark in the 331SUSYRN.}
\end{center}
\label{f8}
\end{figure}

\begin{figure}[t]
%
%
\begin{center}
\begin{picture}(220,100)(0,0)
\SetWidth{1.1}
\ArrowLine(10,10)(50,50)
\Text(10,10)[r]{$u_{1}$}
\ArrowLine(10,90)(50,50)
\Text(10,90)[r]{$d_{1}$}
\Vertex(50,50){2.0}
\Text(30,50)[]{$\lambda^{\prime \prime}$}
\DashLine(90,50)(50,50){3}
\Vertex(90,50){2.0}
\Text(80,40)[]{$g$}
\Text(70,55)[b]{$\tilde{d}^{c}_{2}$}
\ArrowLine(90,90)(90,50)
\Text(95,90)[l]{$d^{c}_{2}$}
\Line(90,50)(130,50)
\Text(110,45)[t]{$\tilde{\gamma}, \tilde{Z}, \tilde{Z}^{\prime}$}
\Vertex(130,50){2.0}
\Text(140,60)[]{$g$}
\ArrowLine(130,50)(130,90)
\Text(160,90)[b]{$d^{\prime}_{1},d^{\prime}_{2}$}
\DashLine(170,50)(130,50){3}
\Vertex(170,50){2.0}
\Text(190,50)[]{$\zeta$}
\Text(160,30)[b]{$\tilde{d}^{\prime c}_{1},\tilde{d}^{\prime c}_{2}$}
\ArrowLine(210,90)(170,50)
\Text(220,90)[l]{$u^{c}_{1},u^{c}_{2}$}
\ArrowLine(210,10)(170,50)
\Text(220,10)[l]{$d^{c}_{1},d^{c}_{2}$}
\end{picture}\\ 
{\sl Figure 9: Proton Decay in exotic quark in the 331SUSYRN.}
\end{center}
\label{f9}
\end{figure}

The baryon-parity forbid the terms in the last two lines in the 
Eq.(\ref{sp3susy2}). In these lines only the term proportional to $\lambda^{\prime}_{\alpha ai}$ and $\xi_{2 \alpha a \beta}$ as in 
the previous case. Therefore the proton is stable at tree-level and also is 
possible to have Leptogenesis and to produce massive neutrinos. 

\section{Conclusions}
\label{concl}
On this article we have studied the proton decay in two differents 331 
supersymmetric model. We showed that in the minimal supersymmetric 331 \cite{331susy} there 
are two differnts contributions, they are $p \to \pi^{+}\nu^{c}_{e}$, $p \to e^{+}\pi^{0}$ and 
$p \to K^{+} e^{\mp} \mu^{\pm}\nu^{c}_{\tau}$ with limits given in 
Eq.(\ref{proton}) and Eq.(\ref{proton2}), respectivelly. On the supersymmmetric 331 model 
with right-handed neutrinos \cite{331susy2} only the modes in $\pi$ are possible 
if we negligible intergenerational mixing. Also we can forbid the proton decay 
imposing baryon-symmetry but at the same time allow Leptogenesis and neutrino 
masses. Of course, if we want to allow proton to decay we have to take in 
account the intergenerational mixings coming from the neutrino's sector and 
from the $d$-quark and $\tilde{d}$-squarks sectors  to suppress 
decays presented on this article.

\begin{center}
{\Large {\bf Acknowledgments}}
\end{center} 
This work was supported by Conselho Nacional de Ci\^encia e Tecnologia (CNPq). 
The author thanks V. Pleitez and J. C. Montero for discussions.

\end{document}